\documentclass[aps,prl,twocolumn,unsortedaddress]{revtex4-1}
\usepackage{epsfig,pstricks,graphicx}
\usepackage{amsmath,amssymb}
\usepackage[mathscr]{eucal}
\usepackage[normalem]{ulem}
\usepackage{color}

\usepackage{accents}

\def\CC{{\rm\kern.24em \vrule width.04em height1.46ex depth-.07ex
\kern-.30em C}}
\def\RR{{\rm
         \vrule width.04em height1.58ex depth-.0ex
         \kern-.04em R}}
\def\P{{\rm I\kern-.25em P}}
\def\id{{\rm 1\kern-.22em l}}

\newcommand{\bra}[1]{\left\langle #1 \right |}
\newcommand{\ket}[1]{\left | #1 \right\rangle}

\newcommand{\tr}{\operatorname{tr}}

\begin{document}

\title{
      Monogamy equalities for qubit entanglement from Lorentz invariance 
      }

\author{Christopher Eltschka}
\affiliation{Institut f\"ur Theoretische Physik, 
         Universit\"at Regensburg, D-93040 Regensburg, Germany}
\author{Jens Siewert}
\affiliation{Departamento de Qu\'{\i}mica F\'{\i}sica, Universidad del Pa\'{\i}s Vasco -
             Euskal Herriko Unibertsitatea, 48080 Bilbao, Spain}
\affiliation{IKERBASQUE, Basque Foundation for Science, 48011 Bilbao, Spain}

\begin{abstract}            
A striking result from nonrelativistic
quantum mechanics is the monogamy of entanglement,
which states that a particle can be maximally 
entangled only with one other party, not with several ones.
While there is the exact quantitative relation 
for three qubits and also several inequalities describing 
monogamy properties
it is not clear to what extent exact monogamy 
relations are a general feature of quantum mechanics.
We prove that in all many-qubit systems 
there exist strict monogamy 
laws for quantum correlations.
They come about through the curious relation
between the nonrelativistic quantum mechanics of qubits and
Minkowski space.
We elucidate the origin of entanglement monogamy 
from this symmetry perspective 
and provide
recipes to construct new families of such equalities.
\end{abstract}

\maketitle

{\em Introduction.--} 
Monogamy of entanglement as a qualitative concept has been discussed
for almost two  decades~\cite{Bennett1996,Terhal2004}. 
Conceivably, this intuition can be
cast into a mathematical framework of inequalities for certain types
of quantum correlations. An influential result is the strong subadditivity
of the von Neumann entropy~\cite{Lieb1973}, among other important examples for
monogamy inequalities~\cite{KW2004,Osborne2006,Fei2008,Adesso2014}. 
Applications of entanglement monogamy pervade many areas of physics, such
as quantum information and the foundations of quantum 
mechanics~\cite{Toner2009,Seevinck2010,Bennett2014}, 
condensed-matter physics~\cite{Walther2011,Brandao2013,Latorre2013}, 
statistical mechanics~\cite{Bennett2014},
and even black-hole physics~\cite{Preskill2013,Susskind2013}. 

\begin{figure}[ht]
  \centering
  \includegraphics[width=.96\linewidth]{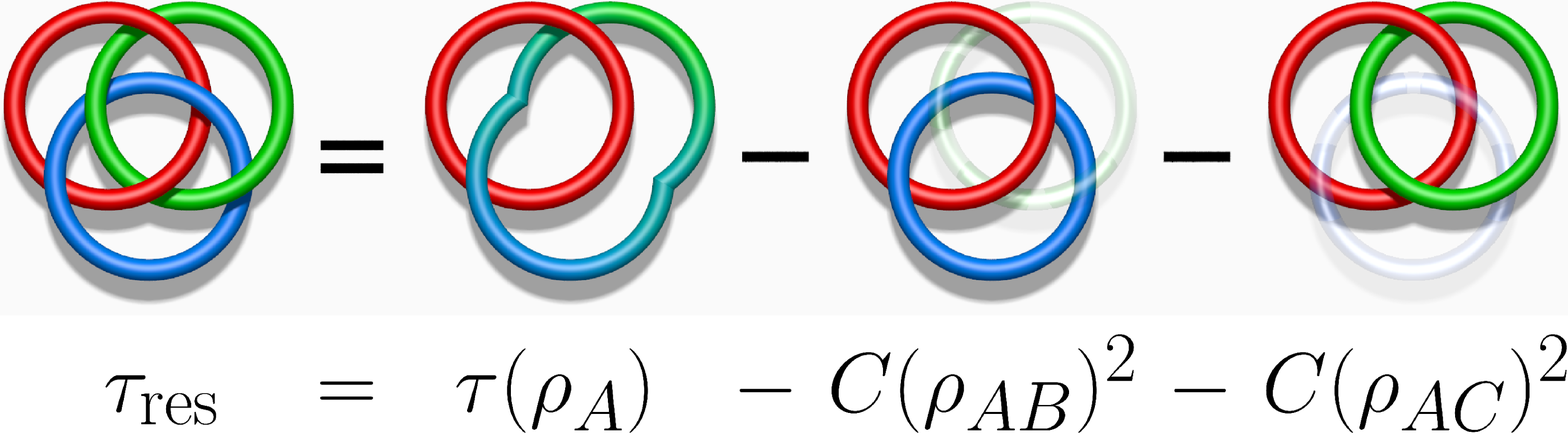}
  \caption{Qualitative sketch of the 
           CKW monogamy equality
           for pure states of three qubits 
           $A$, $B$, $C$.
           The Borromean rings on the left-hand side
           adequately illustrate the tripartite entanglement quantified
           by $\tau_{\mathrm{res}}$. 
           The first term on the right-hand side
           stands for the linear entropy $\tau(\rho_A)\equiv\tau(\rho_{A(BC)})$
           of qubit $A$ that measures the
           entanglement between $A$ and the composite system $\{BC\}$.
           The remaining two terms represent the amount of
           bipartite entanglement
           of $A$ with $B$ or $C$ quantified by the concurrences
           $C(\rho_{AB})^2$
           and $C(\rho_{AC})^2$, respectively, thereby 'forgetting'
           (tracing out) the third party.
           While $\tau_{\mathrm{res}}$ is a global property of the
           state, the quantities on the right-hand side refer to
           different reduced states.
  \label{fig:ckw}
          }
\end{figure}

This has to be contrasted with the possibility of monogamy {\em equalities},
i.e., exact relations for different types of correlations in arbitrary
pure quantum states. Although several such equalities are known 
(see below), the only widely recognized---and at the same time 
perhaps most famous---result is  the three-qubit
monogamy relation discovered by Coffman, Kundu, and Wootters 
(CKW)~\cite{CKW2000}, cf.\ Fig.~\ref{fig:ckw}.

The existence of monogamy {\em inequalities}
appears not entirely unexpected,
because intuitionally one would associate them with convexity  properties
of the quantum-mechanical state space. As opposed to this,
rigorous monogamy expressed in terms of {\em equalities} for pure states 
represent a much stronger constraint and
hint at a certain fine 
tuning of the mathematical properties of quantum states.
In fact, one might even consider using different names for equalities
vs.\ inequalities, instead of terming them both `monogamy relations'.

In this article, we show for qubit systems 
that such monogamy equalities are not coincidental, but
represent a universal feature of single-copy entanglement
that is deeply rooted in the 
algebraic structure of quantum theory.  Since exact monogamy relations 
link those properties to local SL(2,$\mathbb{C}$) 
invariants~\cite{Brylinski2002,Verstraete2003} they confirm
the central importance of these quantities for entanglement theory.

To be slightly more specific, the quantum-mechanical state $\rho_{ABC\cdots}$
shared between several parties $A$, $B$, $C$, $\ldots$ contains all
available information about the correlations between the
individual parties. On the other hand, 
the reduced state, for example $\rho_{AC(B\cdots) }\equiv \rho_{AC}$
describes the state of the subset $\{AC\}$ of the partners
and `forgets' about the information regarding the other
parties. Technically, the reduced state is obtained by 
tracing out the degrees of freedom of the other parties. 
For a multipartite system there are many ways to form subsets of parties,
and the corresponding reduced states.
We may quantify the entanglement contained in each
reduced state by an appropriate entanglement measure. A monogamy
relation is nothing but a mathematical constraint 
for the entanglement
quantifiers of different reduced states of a composite quantum system.
%
%

%
\begin{figure}[h]
  \centering
  \vspace*{5mm}
  \includegraphics[width=1.\linewidth]{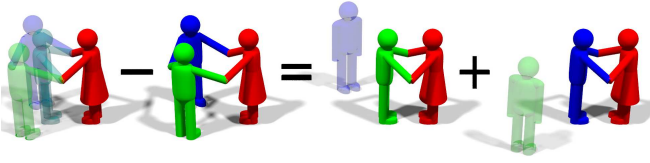}
  \caption{Why is the CKW equality in Fig.~\ref{fig:ckw} termed a 
           'monogamy relation'? 
           The linear entropy $\tau(\rho_A)$ can be viewed as the 
           total amount of Alice' social connectivity (first picture on 
           the left), while the
           residual tangle $\tau_{\mathrm{res}}$ represents 
           connectivity in a group that Alice, Bob, and Charlie
           share symmetrically (2nd picture on the left). 
           On the other hand, the concurrences $C(\rho_{AB})^2$,
           $C(\rho_{AC})^2$ characterize Alice' personal relation 
           with Bob or Charlie, respectively. As the equality shows, if Alice'
           total connectivity and the group connectivity are fixed,
           so is the sum of her personal relations. That is, Alice may
           share relations with Bob and Charlie, but she cannot
           dedicate maximal attention to both Bob and Charlie at the same time.
          }
  \label{fig:monog}
\end{figure}
%

{\em Bloch representation and Lorentz invariance.--}
Usually, nonrelativistic quantum mechanics is constructed starting
from pure states in Hilbert space, which later are generalized to mixed states
representing ensembles of pure states. In our work, we emphasize
the opposite point of view, that is, that states are positive Hermitian 
operators (or 'density matrices') and pure states are 
special states characterized by additional conditions. 
It is those
conditions that imply peculiar properties such as strict monogamy
of entanglement. In order to see this, it is essential to choose
a representation for the density matrices that 
adequately encodes the relevant
state properties, 
the Bloch representation~\cite{Eberly1981,Teodorescu2003,Mahler2004,Bengtsson2006}.

Consider the state $\rho$ of a single qubit which can be 
written as~\cite{Eberly1981,Teodorescu2003,Mahler2004,Bengtsson2006}
\begin{equation}
      \rho\ =\ \frac12 \left( r_0 \id_2 + r_1 \sigma_1 + r_2 \sigma_2
                                        + r_3 \sigma_3 \right)
\label{eq:bloch1}
\end{equation}
where $\sigma_j$ are the Pauli matrices, $\id \equiv \sigma_0$ is
the identity operator for qubits, and $r_j$ are real numbers. Usually,
$\rho$ is normalized so that $r_0=1$. An important quantity
is the determinant~\cite{Teodorescu2003} 
\begin{eqnarray}
     4\det\rho\ =\ 2\tr \rho\sigma_2\rho^T\sigma_2
              &\ =\ &  r_0^2  - r_1^2 -r_2^2 - r_3^2 
\nonumber\\
              &\ \equiv\ & r_{\mu}r^{\mu}
\label{eq:minkowski1}
\end{eqnarray}
because it does not change under determinant-one operations 
$F\in \mathrm{SL}(2,\CC): \rho\rightarrow F\rho F^{\dagger}$. 
We have introduced the Einstein summation convention 
$r_{\mu}r^{\mu}\equiv \sum_{\mu,\nu} \eta^{\mu\nu} r_{\mu} r_{\nu}$
and the Minkowski metric $\eta^{\mu\nu}=\mathrm{diag}(1,-1,-1,-1)$.
As is known from Lie theory, 
local SL invariance translates into
Lorentz invariance in the Bloch coefficients $r_{\mu}$.
This 
representation can be generalized to any number $N$ of qubits
\begin{equation}
      \rho\ =\ \frac{1}{2^N} \sum_{j_1,\ldots,j_N} r_{j_1j_2\ldots j_N}
                             \sigma_{j_1}\otimes\ldots\otimes\sigma_{j_N}
\label{eq:blochN}
\end{equation}
as well as 
\begin{equation}
     \tr R\ \equiv\
 \tr \rho\sigma_2^{\otimes N}\rho^T\sigma_2^{\otimes N}
     \ =\ \frac{1}{2^N}\ r_{\mu_1\ldots\mu_N}r^{\mu_1\ldots\mu_N}
\label{eq:minkowskiN}
\end{equation}
(using the transposed density matrix $\rho^T$).
This expression is invariant under local determinant-one operations 
implying Lorentz invariance independently on each qubit 
index~\cite{Teodorescu2003}. 
It is also nonnegative since $R$ can be rearranged under the trace
as a positive operator~\cite{Wootters1998} $\tr R=
 \tr \sqrt{\rho}\sigma_2^{\otimes N}\rho^T\sigma_2^{\otimes N}
     \sqrt{\rho}$.
The aforementioned
relations \eqref{eq:minkowski1} and \eqref{eq:minkowskiN} are
valid for arbitrary states, but we may ask whether for pure states
more elaborate predicitions are possible.

{\em Origin of exact monogamy.--}
The space of all states $\rho$ is a convex set, with the pure
states $\pi_{\psi}$ as extreme points. 
They can be characterized as  projectors
\begin{equation}
       \pi_{\psi}\ =\ \pi_{\psi}^2 \ \ .
\label{eq:pure2}
\end{equation}
Here, $\psi$ refers to the the usual bra-ket notation,
that is, $\pi_{\psi}=\ket{\psi}\!\bra{\psi}$. Inserting 
Eq.~\eqref{eq:blochN} into \eqref{eq:pure2} yields an operator identity
that has to be satisfied termwise. The most prominent among these conditions is
the normalization of $\pi_{\psi}^2$
\begin{eqnarray}
     \tr \pi_{\psi}^2
      \ =\  1
    \ & = &\
\frac{1}{2^N} \sum r_{\mu_1\ldots\mu_N}^2
\ \ .
\label{eq:euclidN}
\end{eqnarray}

Surprisingly, this is enough to explain the origin
of monogamy equalities. To this end, we write the
Minkowskian and Euclidean sums \eqref{eq:minkowskiN} and \eqref{eq:euclidN}
explicitly separating time-like and space-like indices (we show an
example for two qubits)
\begin{eqnarray}
     2^2\tr R_{\psi}
              &\ =\ & r_{00}^2-\sum_{j=1}^3 \left(r_{0j}^2+r_{j0}^2 \right)
                      + \sum_{j,k=1}^3 r_{jk}^2
\label{eq:s}
\\
              &\ \equiv\ & S_0 - S_1+ S_2
\label{eq:sM}
\\
     2^2\tr \pi_{\psi}^2
                  &\ =\ & S_0 + S_1+ S_2
\ \ .
\label{eq:sE}
\end{eqnarray}
The symbol $S_k$ denotes 
the sum of all terms $r_{\ldots}^2$
with $k$ space-like indices. 
Each quantity $S_k$ is invariant under local unitaries.
The coefficients 
$r_{\ldots}$ with a time-like index 0 at position $q$ are 
components of a reduced state $\tr_q \pi_{\psi}$. Finally,
for pure states we have $\tr R_{\psi}=|H(\psi)|^2$
with the well-known polynomial 
invariant~\cite{Wong2001,Brylinski2002,Eltschka2012}
 $H(\psi)$, i.e.,
a quantity that characterizes and quantifies 
global entanglement in the pure state $\pi_{\psi}$.

It is evident from
Eqs.~\eqref{eq:sM} and \eqref{eq:sE} that the term $S_2$ with {\em only}
space-like indices can be eliminated so that we are left with an
equation that contains only quantities that either characterize 
global entanglement in the state or describe elements of reduced states.
This {\em is} the general monogamy principle that obviously works for
any number of qubits since the corresponding relations for purity and 
$|H(\psi)|^2$ are always independent.
The 
question is whether and how the terms of the reduced states
can be related to entanglement measures in a simple manner.

{\em Degree-2 monogamy relations.--}
In the following we show how specific monogamy relations can be
deduced from this principle, thereby focusing on expressions that
are quadratic in the density matrix.
By either subtracting (even $N$) or adding (odd $N$)
the equations for $\tr R$ and $\tr \rho^2$ 
the term with only space-like indices is eliminated 
and we obtain an equality for all $N$-qubit mixed states
\begin{equation}
        (-1)^N\tr R\ =\  \tr\rho^2 - \frac{1}{2^{N-1}} 
           \sum_{k=a_N}^{\lfloor N/2\rfloor } \ S_{2k-a_N}
\label{eq:simple-mixed}
\end{equation}
where $\lfloor N/2\rfloor $ denotes 
the largest integer not exceeding $N/2$, and $a_N=\frac12\left[1+(-1)^N\right]$.

For pure states we use Eq.~\eqref{eq:euclidN}. 
The sums $S_k$ can be expressed in terms of the purities
$\tr\rho_{\{j\}}^2$ 
of the reduced density matrices $\rho_{\{j\}}$ with a set of $j$ qubits
traced out. The purities are related to the linear entropy
$\tau(\rho)=2\left[(\tr\rho)^2-\tr\rho^2\right]$,
and we can straightforwardly derive a monogamy relation
for all integers $N\geqq 2$
\begin{align}
   2 |H(\psi)|^2  = 
                  \tau_{(1)}-  &  \tau_{(2)}+\tau_{(3)}
                                   -+\ldots +
                                   (-1)^N\tau_{(N-1)}  
\ \ .
\label{eq:evenMonog}
\end{align}
Here, $\tau_{(j)}\equiv \sum_{\{j\}} \tau(\tr_{\{j\}}\pi_{\psi})$
is the sum of all linear entropies that are obtained by tracing out
$j$ qubits from the pure state $\psi$. 
We emphasize that we use linear entropies in Eq.~\eqref{eq:evenMonog}
just for the sake of a transparent notation. As we are dealing with pure states
$\psi$ here, we can equally well substitute (for each bipartition
$A|B$)
\begin{equation}
    \tau(\tr_B\pi_{\psi} )\ =\ C_{A|B}(\psi)^2
\label{eq:linent-conc}
\end{equation}
where $C_{A|B}(\psi)$ is the concurrence of the state $\psi$ 
with respect to this bipartition. That is, the right-hand side of 
Eq.~\eqref{eq:evenMonog} is a combination of entanglement monotones
on the respective bipartitions.
For $N=4$ this relation was found in Ref.~\cite{Gour2010}.
%
\begin{figure}[ht]
  \centering
  \includegraphics[width=1.\linewidth]{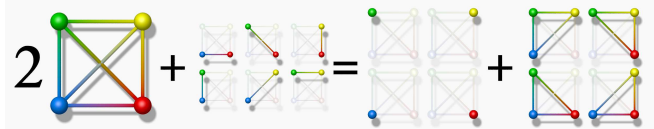}
  \caption{Illustration of the degree-2 monogamy relation 
           for pure four-qubit states. The big square on the left represents 
           $|H(\psi)|^2$ while the other items stand for the 
           six possibilities of two-qubit linear entropies.
           Correspondingly, on the right-hand side 
           the sums of all single-qubit and three-qubit
           linear entropies are shown. Note that 
           the terms on the right are pairwise identical
           according to equation~\eqref{eq:purities-in-pure}.
           For example, the single-qubit linear entropy of
           the 'green qubit' equals that of the other three
           qubits (lower right entry of the three-qubit 
           contributions).
  \label{fig:4bit}
          }
\end{figure}
%
%
Note that for odd qubit number $\tr R_{\psi}=|H(\psi)|^2\equiv 0$
which leads to an identity 
because 
\begin{equation}
   \tr_B(\tr_A\pi_{\psi})^2\ =\ \tr_A(\tr_B\pi_{\psi})^2
\label{eq:purities-in-pure}
\end{equation}
for any bipartition $A|B$ in a pure state.

Thus, we have found the simplest monogamy relations which
contain, in a sense, averaged quantities (cf.~Fig.~\ref{fig:4bit}). 
For example, for two qubits
$A$ and $B$
the invariant polynomial equals Wootters' concurrence~\cite{Wootters1998}
 $|H(\psi)|=C(\psi)$ so that 
$C(\psi)^2= \frac12 \tau_{(1)}=\frac12[\tau(\rho_A)+\tau(\rho_B)]$
while it is known that $ C(\psi)^2= \tau(\rho_A)=\tau(\rho_B)$.
In order to obtain these sharper relations one can use 
Eq.~\eqref{eq:purities-in-pure}.
Nonetheless Eq.~\eqref{eq:evenMonog} is remarkable: On the 
right-hand side, all terms are invariant under local unitaries
while   
the left-hand side  is SL(2,$\mathbb{C})$ invariant
on all parties. This is one of the hallmarks of entanglement monogamy.
Notably, the relation is homogeneous in the state.

{\em Degree-4 monogamy relations.--}
Our strategy for obtaining monogamy relations so far was:
Consider a local SL invariant expression for the state, reveal
its Lorentz-invariant structure and rewrite it
in terms 
of the reduced states, thereby eliminating
the exclusively space-like parts. 
Until now we have considered only the invariant $H(\psi)$ which is
of homogeneous degree 2 in the coefficients of the Hilbert vector $\psi$.
In the next step we investigate degree-4
invariants which will lead us also to the CKW monogamy
relation.

A local SL invariant of degree 4 
deriving from the $\mathcal{B}^{(N)}$ invariants for odd $N$ qubits
in Ref.~\cite{Eltschka2012} is
\begin{equation}
 \left(\tr \rho\Sigma_{\gamma_1}\rho^T\Sigma_{\gamma_2}\right)
     \left(\tr\rho\Sigma^{\gamma_1}\rho^T\Sigma^{\gamma_2}\right)
     \ \equiv\ B_C^{(3)}(\rho)
\label{eq:T3}
\end{equation}
with
$\Sigma_{\gamma}\equiv\sigma_2\otimes\sigma_2\otimes\sigma_{\gamma}$.
Pure states satisfy
$B_C^{(3)}(\psi)=\left|\mathcal{B}_C^{(3)}(\psi)\right|^2$.
Here we define also the $\mathcal{B}^{(N)}$ invariants with
the full Minkowski metric (as opposed to~\cite{Eltschka2012}).
We obtain
\begin{equation}
   B_C^{(3)}(\psi) 
                   \ = \ \frac{1}{2^4}
                                      \ r_{\alpha\beta\gamma}
                                        r^{\alpha\beta\nu}
                                        r_{\lambda\mu\nu}
                                        r^{\lambda\mu\gamma}
                   \  =\  \tau_{\mathrm{res}}(\psi)^2 
\ \ .
\label{eq:3res-tangle-simple}
\end{equation}
In order to eliminate the space-like indices on qubit $C$ we 
use the following pure-state identities which can be verified 
using the Schmidt decomposition.
Abbreviating $\tilde{\pi}_{\psi}\equiv \Sigma_2\pi_{\psi}^T\Sigma_2$
we have
\begin{eqnarray}
   \left(\tr R_{AB}\right)^2 & = &   \tr\big[
                    \pi_{\psi}\left(\tr_C\tilde{\pi}_{\psi}
            \tr_{AB}\tilde{\pi}_{\psi}     \right)\pi_{\psi}
                                        \big]
\nonumber\\
   \tr R_{AB}^2 & = &            \tr \big[    \left(\tr_C\pi_{\psi}\right)
                                                           \tilde{\pi}_{\psi}
                                 \tr_{AB}\left(\tr_C\pi_{\psi}
                                              \right)
                                                \tilde{\pi}_{\psi}
                                     \big]
\nonumber
\label{eq:Rreduced}
\end{eqnarray}
where the right-most trace is taken first and
traces have to be read such that their argument 
extends all the way
to the right, except for the $C$ traces which remain within their parentheses.
Thus,
\begin{eqnarray*}
       \left|\mathcal{B}_C^{(3)}(\psi)\right|^2
                   \ & = & \  8\left[\left(\tr R_{AB}\right)^2- \tr R_{AB}^2
                           \right]
\\
                     & = &\ 2\Big[ \tr R_{AB} - C(\rho_{AB})^2\Big]^2
\end{eqnarray*}
where $C(\rho_{AB})$ is the concurrence of the rank-2 state
$\rho_{AB}$ and the calculation follows the spirit 
     of Ref.~\cite{CKW2000}.
For $N>3$ all derivations are completely analogous and
we obtain
\begin{equation}
       \left|\mathcal{B}_j^{(N)}(\psi)\right|  
                    \ = \
                           2\Big( \tr R_{[j]} - |H(\tr_j\pi_{\psi})|^2 
                        \Big)
\label{eq:tau3R}
\end{equation}
where $\tr R_{[j]}\equiv \tr R(\tr_j\pi_{\psi})$
and $|H(\tr_j\pi_{\psi})|$ is  the convex roof of $|H|$ 
for the rank-2 state $\tr_j\pi_{\psi}$ 
according to the Wootters-Uhlmann 
method~\cite{Wootters1998,Uhlmann2000}.
Further, $\tr R_{[j]}$ may be replaced using equation~\eqref{eq:minkowskiN}.
The simplest way to get a quartic monogamy
 equality like~\eqref{eq:evenMonog} 
is to add the 
relations~\eqref{eq:tau3R} for the $\left|\mathcal{B}^{(N)}_j\right|$
of all qubits 
\begin{eqnarray}
 \sum_{j=1}^N  \Big[\  &&
     \left|\mathcal{B}_{j}^{(N)}(\psi)\right|\  +\
 2 \left|H(\tr_j\pi_{\psi})\right|^2 
 \ \Big]
\ \ \ \ \ \
\nonumber\\
  = &&\ (-1)^N \sum_{j=1}^{N-1} (-1)^{j+1} j\ \tau_{(j)}
                \ \ .
\label{eq:simple-odd}
\end{eqnarray}
We mention that CKW is obtained by adding the relations for
$\mathcal{B}^{(3)}_C$ and $\mathcal{B}^{(3)}_B$ only.

Curiously, for even $N$ the quartic monogamy equality is not 
strictly of degree 4 
because $\left|\mathcal{B}_j^{(2m)}(\psi)\right|= \left|H(\psi)\right|^2$,
and by using~\eqref{eq:purities-in-pure}
we get back to Eq.~\eqref{eq:evenMonog}, since all
terms $|H(\tr_j\pi_{\psi})|$ vanish.
Yet there do exist  degree-4 monogamy 
equalities also for $N=4$ (cf.~Ref.~\cite{Eltschka2012}), e.g.,
\[
\left|\mathcal{B}_{12}^{(4)}(\psi)-\mathcal{B}^{(4)}_{13}(\psi)\right|^2\
=\ 48^2 \det (\tr_{14}\pi_{\psi})\ \ .
\]

{\em Discussion.--}
We have derived 
degree-2 and degree-4 monogamy relations
of pure-state qubit entanglement, the central results being 
Eqs.~\eqref{eq:evenMonog} and 
\eqref{eq:simple-odd}. 
They can be interpreted as follows.
   Tracing out $B$ in a bipartition $A|B$
   removes the coherences between $A$ and $B$,
   thus converting their mutual quantum correlations into classical
   correlations of $A$ alone.
The global accounting of all these
correlations, which are quantified by the linear entropies, 
is given by the SL$(2,\mathbb{C})^{\otimes N}$ invariants on the left-hand side
of Equations~\eqref{eq:evenMonog} and \eqref{eq:simple-odd}. 
The right-hand sides can be viewed as decompositions 
of the global into bipartite correlations. 
We mention that for qubit systems, there are algebraically independent
local SL invariants also of degree 6 (for $N\geqq 4$) as well as of degree 8, 
10, etc.~($N>4$). We expect that it is
possible to find corresponding monogamy equalities also for those invariants
by continuing the hierarchy described in this article.

%
%
%
%
%
%
%
%
%
%
%
%

\vspace*{5mm}
This work was funded by the German Research Foundation within 
SPP 1386 (CE), by Basque Government grant IT-472,
MINECO grant FIS2012-36673-C03-01, and UPV/EHU program
UFI 11/55 (JS).
The authors thank A.\ Uhlmann for helpful remarks, 
and J.\ Fabian, J.G.\ Muga, and K.\ Richter for their support.
%
%
%
%

%

\end{document}